# Implementation of an efficient Fuzzy Logic based Information Retrieval System

**Narina Thakur**
Professor, Bharati Vidyapeeth's College of Engineering/Computer Science, New Delhi, 110058, INDIA
Email: narina.thakur@bharatividyapeeth.edu

**Prabhjot Singh, Sumit Dhawan and Shubham Agarwal**
Student, Bharati Vidyapeeth's College of Engineering/Computer Science, New Delhi, 110058, INDIA
Email: mail@prabh.in, sdhawan999@gmail.com, shubhamagar05@gmail.com

*Abstract*— This paper exemplifies the implementation of an efficient Information Retrieval (IR) System to compute the similarity between a dataset and a query using Fuzzy Logic. TREC dataset has been used for the same purpose. The dataset is parsed to generate keywords index which is used for the similarity comparison with the user query. Each query is assigned a score value based on its fuzzy similarity with the index keywords. The relevant documents are retrieved based on the score value. The performance and accuracy of the proposed fuzzy similarity model is compared with Cosine similarity model using Precision-Recall curves. The results prove the dominance of Fuzzy Similarity based IR system.

## I. INTRODUCTION

An information retrieval system stores and indexes documents such that when users express their information need in a query the system retrieves the related documents associating a score to each one. The higher the score the greater is the importance of the document. Usually an information retrieval system returns large result sets and the users must spend considerable time until they find the items that are actually relevant. Moreover, documents are retrieved when they contain the index terms specified in the queries. However, this approach will neglect other relevant documents that do not contain the index terms specified in the user's queries. When working with specific domain knowledge this problem can be overcome by incorporating a knowledge base which depicts the relationships between index terms into the existing information retrieval systems [1].

To deal with the vagueness typical of human knowledge, the fuzzy set theory can be used to manipulate the knowledge in the bases. The expectation is that the indexed terms can improve the quality of retrieved documents bringing the most relevant and more semantically related to the initial query. Full-text search is still the most popular form of search and is very useful to retrieve documents for which we know the keywords to search for. Indeed, full-text search is not suitable for finding relevant documents about a specific topic in the context of a given task [2].

Another problem relies in the huge number of retrieved documents/data. It is difficult for a user to deal with thousands of electronic documents. The techniques often used by search engines are based on statistical methods and do not permit to take account of the semantics contained in the user's query as well as in the documents. Some approaches have been developed to extract semantics and so, to better answer user queries. On the other hand, most of these techniques have been designed to be applied to the whole web and not on a particular domain [2].

Apache Lucene has been used to retrieve the relevant documents. Lucene is a text search engine library suitable for nearly any application that requires full-text search. Lucene is composed of many text processing tools. Each tool within it is a heuristic, an algorithmic shortcut in lieu of true linguistic comprehension [3].

Lucene performs a fuzzy search in two steps. First, Lucene searches for tokens stored in the database that are similar to the query tokens. To determine if tokens are similar, Lucene computes an edit distance (also referred to as a Levenshtein Distance) from the query tokens to the tokens stored in the database. Second, Lucene uses the similar tokens it finds as new query tokens to retrieve relevant documents [4].

In this paper, we propose and discuss the implementation and performance evaluation details of an IR system with fuzzy based similarity measures. Experiments have been performed on the TREC Ohsumed [5] data collection. Ohsumed.87 (60,303,307) contains the MEDLINE documents for the year 1987. This test collection was created to assist information retrieval research. The relevance of documents using fuzzy similarity is compared with the QRELs provided on Ohsumed website [5]. Precision and Recall values are calculated using trec_eval script. The description of the implementation and experimental results will be discussed briefly.

## II. BOOLEAN INFORMATION RETRIEVAL MODEL

Currently, most of the commercial information retrieval systems are based on the Boolean logic model. Boolean IR model assumes that the user's query appropriately describes the user's need. There is no scope for vagueness or Fuzziness.

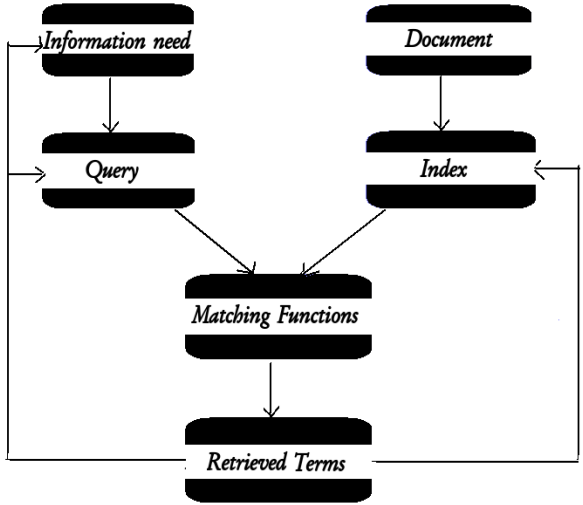

Fig. 1. Visualization of Information Retrieval system process.

### A. Issues and solutions

Traditional IR systems assume that a user's queries can precisely be characterized by the index terms. However, this assumption is inappropriate due to the fact that the user's queries may contain fuzziness. The reason for the fuzziness contained in the user's queries is that the user may not know much about the subject he/she is searching or may not be familiar with the information retrieval system. Therefore, the query specified by the user may not describe the information request properly. Since fuzzy set theory can be used to describe imprecise or fuzzy information, many researchers have applied the fuzzy set theory to information retrieval systems.

Other issues such as performance, scalability and occurrences of paging update are other common information retrieval issues. Relevance is the relational value of a given user query to the documents within the database. Relevance of a document is normally based on a document ranking algorithm. These algorithms define how relevant a document is to a user query by using functions that define relations between the query given and the documents collected in the index. The evaluation of the feedback given by the information retrieval system is another issue with information retrieval. The behavior of the system may not meet the expectations of the user or the documents returned from the system may not all be relevant to a query [6].

Depending on the system and the user, the results of a query should be in a format that most fits the data being searched and returned. Information needs is how the user interacts with the information retrieval system. The data within the system should be able to be accessed easily and in a way that is convenient to the user. Retrieving too much information might be inconvenient in certain systems, also in other systems not returning all relevant information may be unacceptable.

As noted, managing volumes of information from the web can be difficult due to the volume of documents the server contains. This leads to an even larger problem when trying to retrieve the most relevant results to a query. A simple retrieval query can return thousands of documents, many of which are loosely related to the original retrieval criteria. Natural language and word ambiguity can also add distress to information retrieval [7]. To overcome this, an information retrieval system needs to have good query management along with the ability to give weight to documents that are more relevant to the user's query, and present those results first.

## III. FUZZY INFORMATION RETRIEVAL SYSTEM

### A. Fuzzy Logic

Fuzzy Logic is basically a logic approach that allows intermediate truth values to be defined between conventional evaluations of true and false. Notions like rather hard or pretty cool can be formulated mathematically and processed by computers [8].

### B. Fuzzy Sets

The very basic notion of fuzzy systems is a Fuzzy set. Fuzzy sets are sets whose elements have degrees of membership. In classical set theory, an element either belongs or does not belong to the set. However, fuzzy set theory permits the gradual assessment of the membership of elements in a set; this is described with the help of a membership function valued in the real interval [0, 1]. More precisely, the transition from membership to non-membership could be gradual rather than abrupt as in the case of Boolean theory.

Nevertheless, there has not been a formal basis for how to determine the grade of membership. It is a subjective measure that depends on the context, set membership does not mean the same thing at the operational level in each and every context. Dubois and Prade have showed in that degree of membership can have one of the following semantics: a degree of similarity, a degree of preference, or a degree of uncertainty [8].

Fuzzy information retrieval systems utilize the tools defined in fuzzy logic and fuzzy relations to infer the best results to a user query. Unlike Boolean systems, fuzzy systems are most effective when dealing with data that may display a degree of membership. In fuzzy systems, objects described in terms of their properties which characterize the objects are assigned relational membership values to show relevancy from properties to objects or vice versa. These values are different than using probability. In probabilistic systems of the discrete case, count, i.e. the total number of an event, is used to compute the probability that a relationship exists, whereas in fuzzy systems, a membership value is used to determine a weighted relational mapping.

*C. Existing Techniques*

The problem of fuzzy string searching can be formulated as follows: "Find in the text or dictionary of size n all the words that match the given word (or start with the given word), taking into account k possible differences (errors)."

For example, if we requested for 'machine' with two possible errors, find the words 'marine', 'lachine', 'martine', and so on.

Meanwhile, in most cases a metric is understood as a more general concept that does not meet the condition above, this concept can also be called distance. Among the most well-known metrics are Hamming, Levenshtein and Damerau-Levenshtein distances. The Hamming distance is a metric only on a set of words of equal length, and that greatly limits the scope of its application.

- **Levenshtein distance:** The Levenshtein distance, also known as "edit distance", is the most commonly used metric, the algorithms of its computation can be found at every turn. Nevertheless, it is necessary to make some comments about the most popular algorithm of calculation - Wagner-Fischer method.

$$\operatorname{lev}_{a,b}(i,j) = \begin{cases} \max(i,j) & \text{if } \min(i,j) = 0, \\ \min \begin{cases} \operatorname{lev}_{a,b}(i-1,j) + 1 \\ \operatorname{lev}_{a,b}(i,j-1) + 1 \\ \operatorname{lev}_{a,b}(i-1,j-1) + 1_{(a_i \neq b_j)} \end{cases} & \text{otherwise.} \end{cases}$$

Fig. 2: Levenshtein distance formula

The original version of this algorithm has time complexity of O(mn) and consumes O(mn) memory, where m and n are the lengths of the compared strings.

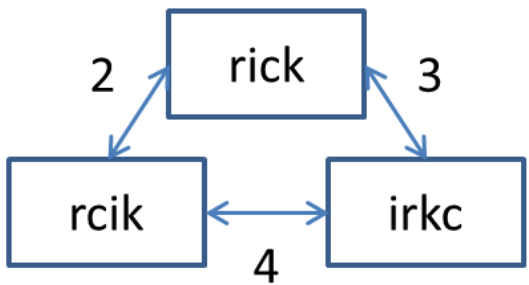

Fig. 3: Example of Levenshtein distance

The entire process can be represented by the following matrix:

| | | E | L | E | P | H | A | N | T |
|---|---|---|---|---|---|---|---|---|---|
| | 0 | 1 | 2 | 3 | 4 | 5 | 6 | 7 | 8 |
| R | 1 | 1 | 2 | 3 | 4 | 5 | 6 | 7 | 8 |
| E | 2 | 1 | 2 | 2 | 3 | 4 | 5 | 6 | 7 |
| L | 3 | 2 | 1 | 2 | 3 | 4 | 5 | 6 | 7 |
| E | 4 | 3 | 2 | 1 | 2 | 3 | 4 | 5 | 6 |
| V | 5 | 4 | 3 | 2 | 2 | 3 | 4 | 5 | 6 |
| A | 6 | 5 | 4 | 3 | 3 | 3 | 3 | 4 | 5 |
| N | 7 | 6 | 5 | 4 | 4 | 4 | 4 | 3 | 4 |
| T | 8 | 7 | 6 | 5 | 5 | 5 | 5 | 4 | 3 |

Fig. 4: Levenshtein distance implementation process

- **Prefix Distance:** Usually it is necessary to calculate the distance between the prefix pattern and a string, to find the distance between the specified prefix and nearest string prefix. In this case, you must take the smallest of the distances from the prefix pattern to all the prefixes of the string. Obviously, the prefix length cannot be considered as a metric in the strict mathematical sense, what limits its application. Often, the specific value of a distance is not as important as fact that it exceeds a certain value.

- **Damerau-Levenshtein distance:** This variation contributes to the definition of the Levenshtein distance one more rule - transposition of two adjacent letters are also counted as one operation, along with insertions, deletions, and substitutions. Therefore, this metric gives the best results in practice [9].

$$d_{a,b}(i,j) = \begin{cases} \max(i,j) & \text{if } \min(i,j) = 0, \\ \min \begin{cases} d_{a,b}(i-1,j) + 1 \\ d_{a,b}(i,j-1) + 1 & \text{if } i,j > 1 \text{ and } a_i = b_{j-1} \\ d_{a,b}(i-1,j-1) + 1_{(a_i \neq b_j)} & \text{and } a_{i-1} = b_j \\ d_{a,b}(i-2,j-2) + 1 \end{cases} \\ \min \begin{cases} d_{a,b}(i-1,j) + 1 \\ d_{a,b}(i,j-1) + 1 & \text{otherwise.} \\ d_{a,b}(i-1,j-1) + 1_{(a_i \neq b_j)} \end{cases} \end{cases}$$

Fig. 5: Damerau-Levenshtein distance implementatation process

To calculate this distance, it suffices to slightly modify the regular Levenshtein algorithm as follows: hold not two, but the last three rows, and add an appropriate additional condition - in the case of transposition take into account its cost.

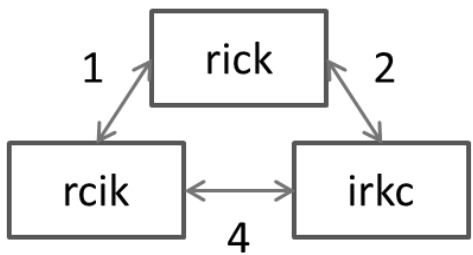

Fig. 6: Example of Damerau-Levenshtein Distance

## IV. DETAILED IMPLEMENTATION

The metric that has been used by fuzzy queries to determine a match is the Levenshtein distance formula. Simply put, the Levenshtein distance between two pieces of text is the number of insertions, deletions, substitutions, and transpositions needed to make one string match the other. For example, the Levenshtein distance between the words "ax" and "axe" is 1 due to the single deletion required. That means that when performing fuzzy queries, the query text may be compared to an unanticipated term value as a result of analysis, leading to sometimes confusing results.

The source code is written in Java in which Lucene packages are imported. The first step is to create an index of all keywords appearing in the Ohsumed dataset followed by searching. A fuzzy retrieval method is implemented similar to the inbuilt FuzzyQuery method with varying parameters.

The parameters specified are:
1) Query terms: The string query specified by the user is converted to query terms which are then searched for in the generated index.
2) Fuzziness: The allowed difference between the keywords and query term that are to be retrieved.
3) Affix Length: The maximum possible character length common to both the query and keyword.

The fuzziness argument specifies that the results match with a maximum edit distance in terms of percentage. It should be noted that fuzziness should only be used with values less than 1. The official documentation still refers to setting fuzziness to float values, like 0.5, but these values are in-fact deprecated, and are harder to reason about [4].

The documents are retrieved on the basis of their score values depending upon the user query. The output documents are displayed in decreasing order on the basis of the score values. The relevance of retrieved documents are compared with the Ohsumed QRELs data to calculate the precision and recall using trec_eval script. Further, the results are compared with the Cosine similarity retrieval results to generate precision-recall curves.

```
Enter your query:
intravascular coagulation
92 document(s) found. Time taken: 516ms
Score: 0.4333313.       File: F:\Lucene\Data\File_000219.txt
Score: 0.27083206.      File: F:\Lucene\Data\File_000222.txt
Score: 0.21704145.      File: F:\Lucene\Data\File_000224.txt
Score: 0.15347148.      File: F:\Lucene\Data\File_000307.txt
Score: 0.15146597.      File: F:\Lucene\Data\87058538.txt
Score: 0.15146597.      File: F:\Lucene\Data\File09.txt
Score: 0.13152444.      File: F:\Lucene\Data\87325763.txt
Score: 0.121470876.     File: F:\Lucene\Data\File_000107.txt
Score: 0.11691066.      File: F:\Lucene\Data\87242958.txt
Score: 0.1102244.       File: F:\Lucene\Data\File_000117.txt
Score: 0.092881575.     File: F:\Lucene\Data\File_000196.txt
Score: 0.08661552.      File: F:\Lucene\Data\File_000200.txt
Score: 0.08252516.      File: F:\Lucene\Data\File_000174.txt
Score: 0.07550464.      File: F:\Lucene\Data\File_000306.txt
Score: 0.073064275.     File: F:\Lucene\Data\File_000177.txt
Score: 0.07045632.      File: F:\Lucene\Data\87312037.txt
Score: 0.06460853.      File: F:\Lucene\Data\File_000243.txt
Score: 0.06101697.      File: F:\Lucene\Data\87198965.txt
Score: 0.05711615.      File: F:\Lucene\Data\File_000151.txt
Score: 0.053931884.     File: F:\Lucene\Data\87057614.txt
Score: 0.053931884.     File: F:\Lucene\Data\87309677.txt
Score: 0.05062041.      File: F:\Lucene\Data\File_0024.txt
Score: 0.043841504.     File: F:\Lucene\Data\File_000232.txt
Score: 0.04285372.      File: F:\Lucene\Data\File_0053.txt
Score: 0.04126258.      File: F:\Lucene\Data\File_000330.txt
Score: 0.03773987.      File: F:\Lucene\Data\87210296.txt
Score: 0.03736511.      File: F:\Lucene\Data\File_000314.txt
Score: 0.03711241.      File: F:\Lucene\Data\File_000291.txt
```

Fig 7: Output Screenshot: Score values of the retrieved documents

## V. PERFORMANCE EVALUATION

Even though Lucene's Levenshtein distance implementation is state of the art, and quite fast, it is still much slower than a plain match query. The runtime of the query grows with the number of unique terms in the index. That is to say, when performing a fuzzy search the main criteria is not how many documents will be returned, but how many unique terms across the cluster there are for the fields being searched. If there are 100 documents with 10,000 unique words apiece, searching that index will be slower than searching 10,000 documents where the field being searched only has 100 unique words.

The primary reason for this slowness is that a standard match query can quickly check the terms index, an internal data structure used by Lucene, for a match and find documents extremely quickly using binary search. This process is fast even for large dictionaries since binary searches scale well. Fuzzy queries, on the other hand, use a more advanced algorithm involving a DFA which must process a large number of terms. Processing the much larger number of terms required for a fuzzy search is always slower than a simple binary search.

The fuzziness setting, which defines the maximum number of characters in the query that may differ, can also have dramatic effects on the performance of a fuzzy query. Till now, various researchers have used the given parameters to analyze and evaluate the performance of IR Systems:

## A. Precision

It is a fraction of documents that are relevant among the entire retrieved documents. Practically it gives accuracy of result.

$$\text{Precision} = |R_a| / |A|$$

- $R_a$: Set of relevant documents retrieved
- A: Set of documents retrieved

## B. Recall

A fraction of the documents that is retrieved and relevant among all relevant documents is defined as recall. Basically, it gives coverage of result.

$$\text{Recall} = |R_a| / |R|$$

- $R_a$: Set of relevant documents retrieved
- R: Set of all relevant documents

## C. Precision-Recall Curve

This curve is based upon the value of precision and recall where the x-axis is recall and y-axis is precision. Instead of using precision and recall on at each rank position, the curve is commonly plotted using 11 standard recall level 0%, 10%, 20%...........100% [10].

Moreover, average similarity value of documents for individual query and average number of retrieved relevant documents can also be used as parameters to check the performance of IR System. If the values for both of these parameters are high then the performance of IR System will be good. Let us now compare the proposed fuzzy logic based similarity algorithm with the conventional cosine similarity measures to highlight the efficiency of the proposed system.

| Query Number | Query | Maximum Score Value |
|---|---|---|
| 1. | menopausal woman | 0.19804347 |
| 2. | intravascular coagulation | 0.4333313 |
| 3. | cancer and hypercalcemia | 0.31233424 |

Table 1: Fuzzy queries on TREC dataset and their respective scores

- **Fuzzy Similarity v/s Cosine Similarity:**

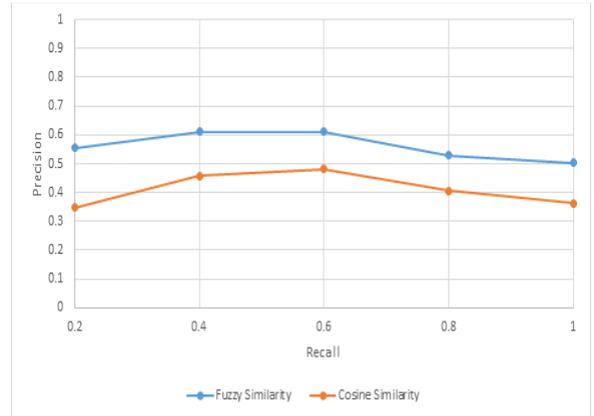

Fig. 6: Comparison of precision recall curve

In Fig. 6, we can clearly see that Fuzzy logic based IR system has high precision and recall values.

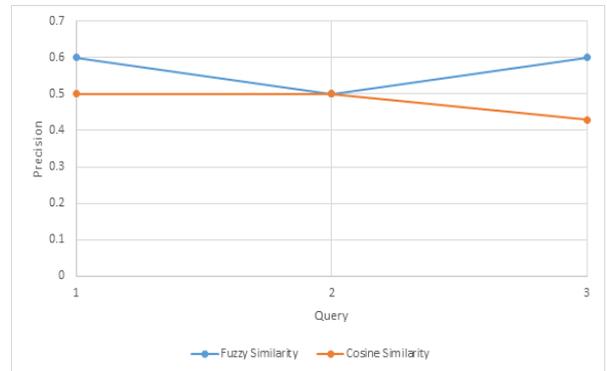

Fig. 7: Comparison of average Precision-Query curve

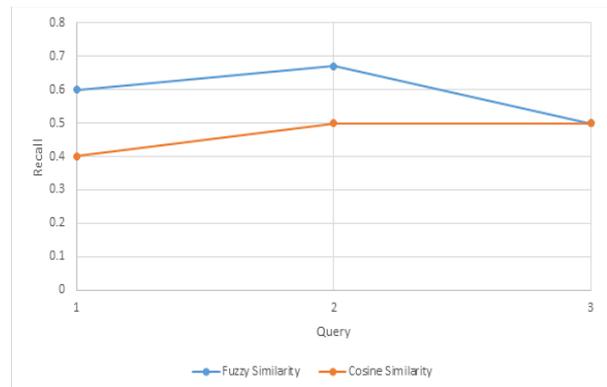

Fig. 8: Comparison of average Recall-Query curve

Fig. 7 and Fig. 8 illustrate that the proposed Fuzzy logic retrieval system gives better average precision and average recall values for all the three queries respectively as compared to cosine similarity measures.

## VI. CONCLUSION

In this paper, we discussed the implementation and efficiency details of an IR system with fuzzy based similarity measures. Experiments performed on TREC Ohsumed data collection using Apache Lucene prove the superiority of the proposed measure. This is a new technique having advantages over the other Information Retrieval systems as it can handle vague and imprecise queries of user very well. The performance of proposed technique is compared with cosine based similarity measure on TREC dataset. Results indicate that proposed similarity measure technique based on fuzzy logic, is better than cosine based similarity measure technique for handling vague, uncertain and imprecise queries. The insight provided by this model makes clear that fuzzy notions describe situations known through imprecise, uncertain, and vague information in a way that neither replaces nor is replaced but that, rather, complements the views produced by other approaches [11].